%% file: main.tex
\documentclass[runningheads]{llncs}

\input{macro.tex}

\title{\tool: Automated Guided Testing for Stateful Deep Learning Systems}

\author{Xiaoning Du\inst{1}\and
Xiaofei Xie\inst{1}\and
Yi Li\inst{1}\and
Lei Ma\inst{2} \and
Jianjun Zhao\inst{3}\and
Yang Liu\inst{1}
}

\authorrunning{Xiaoning Du, Xiaofei Xie, Yi Li, Lei Ma, Jianjun Zhao, and Yang Liu}

\institute{Nanyang Technological University \and Harbin Institute of Technology
\and Kyushu University
}

\begin{document}

\maketitle

\begin{abstract}
Deep learning (DL) defines a data-driven programming paradigm that automatically composes the system decision logic from the training data. In company with the data explosion and hardware acceleration during the past decade, DL achieves tremendous success in many cutting-edge applications. However, even the state-of-the-art DL systems still suffer from quality and reliability issues. It was only until recently that some preliminary progress was made in testing feed-forward DL systems.In contrast to feed-forward DL systems, recurrent neural networks (RNN) follow a very different architectural design, implementing temporal behaviours and ``memory'' with loops and internal states. Such stateful nature of RNN contributes to its success in handling sequential inputs such as audio, natural languages and video processing, but also poses new challenges for quality assurance.

In this paper, we initiate the very first step towards testing RNN-based stateful DL systems. We model RNN as an abstract state transition system, based on which we define a set of test coverage criteria specialized for stateful DL systems. Moreover, we propose an automated testing framework, \tool, which systematically generates tests in large scale to uncover defects of stateful DL systems with coverage guidance. Our in-depth evaluation on a state-of-the-art speech-to-text DL system demonstrates the effectiveness of our technique in improving quality and reliability of stateful DL systems.

\end{abstract}

\input{intro.tex}

\input{background.tex}
\input{overview.tex}
\input{modeling.tex}

\input{criteria.tex}

\input{testing_framework.tex}
\input{evaluation.tex}
\input{relatedwork.tex}
\input{conclusion.tex}

\bibliographystyle{IEEEtran}
\footnotesize
\bibliography{references}
\end{document}

%% file: macro.tex
\usepackage{times}  %Required
\usepackage{helvet}  %Required
\usepackage{courier}  %Required
\usepackage{url}  %Required
\usepackage{graphicx}  %Required
\usepackage{cite}
\usepackage{amsmath,amsfonts}
\usepackage[T1]{fontenc}
\usepackage{xcolor}
\usepackage{graphicx}
\usepackage{booktabs}
\usepackage{microtype}
\usepackage{amsfonts}
\usepackage{multirow}
\usepackage{xspace}
\usepackage{amssymb}
\usepackage[nointegrals]{wasysym}
\usepackage{color}
\usepackage{enumitem,etoolbox}
\usepackage{epsfig}
\usepackage{graphicx}
\usepackage[ruled,vlined,linesnumbered]{algorithm2e}
\usepackage{enumitem}
\usepackage{moresize}
\usepackage{epstopdf}
\usepackage{array}
\usepackage{diagbox}
\usepackage[export]{adjustbox}
\usepackage{tcolorbox}
\usepackage{colortbl}
\usepackage{extarrows}
\usepackage{wrapfig}
\usepackage{subcaption}
\usepackage[hang,flushmargin]{footmisc} % remove footnote indentation

\renewcommand{\paragraph}[1]{\vskip 0.05in \noindent {\bf #1.}}

\newcommand{\yang}[1]{\textbf{\textcolor{blue}{}}}

\makeatletter
\newcommand{\thickhline}{%
    \noalign {\ifnum 0=`}\fi \hrule height 1pt
    \futurelet \reserved@a \@xhline
}
\newcolumntype{"}{@{\hskip\tabcolsep\vrule width 1pt\hskip\tabcolsep}}
\makeatother
\newcommand{\tool}{\emph{DeepCruiser}\xspace}
\renewcommand{\vec}{\mathbf}
\newcommand{\states}{\ensuremath{\mathcal{S}}\xspace}
\newcommand{\inputs}{\ensuremath{\mathcal{X}}\xspace}
\newcommand{\outputs}{\ensuremath{\mathcal{O}}\xspace}
\newcommand{\trans}{\ensuremath{\delta}\xspace}
\newcommand{\atrans}{\ensuremath{\hat{\delta}}\xspace}
\newcommand{\atransfun}{\ensuremath{\hat{\mathcal{T}}}\xspace}
\newcommand{\abss}{\ensuremath{\hat{s}}\xspace}
\newcommand{\absx}{\ensuremath{\hat{x}}\xspace}

\newcommand{\astates}{\ensuremath{\hat{\mathcal{S}}}\xspace}
\newcommand{\ainputs}{\ensuremath{\hat{\mathcal{X}}}\xspace}

\makeatletter
\DeclareRobustCommand\onedot{\futurelet\@let@token\@onedot}
\def\@onedot{\ifx\@let@token.\else.\null\fi\xspace}

\makeatother

%% file: intro.tex
\section{Introduction}
Deep learning (DL) experiences significant progress over the past decades in achieving competitive performance of human intelligence in many cutting-edge applications such as image processing 
\cite{ciregan2012multi}, speech recognition \cite{hinton2012deep}, autonomous driving 
\cite{DBLP:journals/corr/HuvalWTKSPARMCM15}, medical diagnosis \cite{ciresan2012deep} and 
pharmaceutical discovery \cite{CHEN20181241}, which until several years ago were still notoriously difficult to solve programmatically.
DL has been continuously redefining the landscape of industry, in penetrating and reshaping almost 
every aspect of our society and daily life.
For example, Automated Speech Recognition (ASR) currently becomes one of the most effective ways for human computer interaction and communication, and is widely integrated in intelligent assistants on everyday mobile device~(e.g., Apple Siri~\cite{siri}, Amazon Alexa \cite{alexa}, 
Google Assistant \cite{googleassistant}, and Microsoft Cortana \cite{cortana}). 
Although Hidden Markov Models (HMM) were widely used in ASR, DL models are the current 
state-of-the-art solutions in tasks including speech recognition and generation \cite{Graves2013, 
Hannun2014, Oord2016}.

However, the current state-of-the-art DL system still suffers from quality, reliability and security issues, which could potentially lead to accidents and catastrophic events when deployed to safety- and security-critical systems.
With the demanding industry trends for real-world deployment of DL solutions, we have witnessed 
many quality and security issues, such as one pixel attack \cite{one_pixel}, Alexa/Siri 
manipulation with hidden voice command \cite{alexa_siri}, and the Google/Uber autonomous car 
accident~\cite{google_crash, uber_crash}. 
Unfortunately, the quality assurance techniques and tool chains for DL systems are still immature, 
which could potentially hinder future industry scale applications of DL solutions 
towards the goal of Software 2.0~\cite{sw2}.

For traditional software, the software development and quality assurance process are well established over past several decades, but the existing techniques and tool chains could not be directly applied to DL systems. 
This is mainly due to the fundamental differences in programming paradigms, development processes, 
as well as structures and logic representations of the software artifacts (e.g., architectures)~\cite{pei2017deepxplore, ma2018deepgauge}.
To bridge the gap between quality assurance of DL system and its practical applications, some 
recent work on testing and verification of feed-forward DL systems (e.g. Convolution Neural Network 
(CNN)) of image processing started to emerge, ranging from testing criteria 
design~\cite{pei2017deepxplore, ma2018deepgauge, 2018arXiv180808444K}, test 
generation~\cite{Dreossi2017, Zhang:2018:DGM:3238147.3238187}, and metamorphic relation based 
testing oracle \cite{tian2018deeptest}, to abstract interpretation based formal 
analysis~\cite{weiss2017extracting}.

Yet, such practices are hardly applicable to the testing of Recurrent Neural Networks (RNN) due to the 
very different architectural designs, often with loops involved, that introduces internal states and enables the ``memorization'' of what could be observed before or after~\cite{Hannun2014}.
Information flows not only from front neural layers to the rear ones, but also from the current iteration to the subsequent ones.
This makes RNN more suitable to process sequential input streams, such as audios, natural language texts and videos, instead of monolithic data such as images.
Although it is tempting to unroll the network and test the unfolded RNN as if it is a feed-forward neural network~\cite{tian2018deeptest}, such a simple approach ignores the internal states of RNNs and therefore could not precisely reflect the dynamic behaviors for a time sequence.
Furthermore, the unrolled networks would contain different numbers of layers given inputs of various lengths, making the calculation of coverage problematic.

As a typical application of RNN, ASR is faced with the problem of inadequate testing, security threats and attacks~\cite{Carlini2018,Yuan2018, 2015arXiv150704761K, 2017arXiv171103280G}.
Despite the urgent demands, it is unfortunate that
there exists no systematic techniques specialized for RNN-based stateful deep learning systems at the moment.
To address these challenges, in this paper, we propose a coverage-guided automated testing framework for RNN-based stateful DL systems.
Considering the unique features of RNN and the input structures it often processes, we first propose to formalize and model a RNN-based DL system as Markov Decision Process (MDP).
Based on the MDP model, we design a set of testing criteria specialized for RNN-based DL systems to capture its dynamic state transition behaviours.
The proposed testing criteria enable the quantitative evaluation on how extensive the RNN's internal behaviors are covered by test data.

Based on this, we further propose an automated testing framework for RNN-based DL system guided by proposed coverage. 
In this paper, we focus on the ASR domain\footnote{Although we focus on testing RNNs of ASR domain due to the currently urgent industry demands, 
the techniques and testing framework proposed in this paper can be generalized to other RNN-based stateful DL systems.}, 
and incorporate 8 metamorphic transformations to generate new audio test inputs. 
Due to the huge test input generation space, we leverage the coverage feedback to guide the testing direction towards 
systematically cover the major functional behaviors and corner cases of a RNN.
We implement our testing framework, \tool, and perform in-depth evaluation 
on a state-of-the-art practical RNN based ASR system to demonstrate the usefulness.

\begin{comment}
The testing criteria intends to capture the internal state space touchable by the representative data~(e.g., training data), as well as transitions across the space.
Concrete state vectors are available with the training data at hand.
Specifically, we first perform abstraction over the discrete representations with space quantization techniques, and get a set of abstracted states.
All concrete state vectors can be mapped to the abstracted space and transitions over abstracted states are build by checking consequent state vectors.
With the MDP model, we define two state-level coverage criteria and three transition-level criteria.
They can be used to examine whether the testing data is adequate enough to reflect actual model quality, and provide guidance for test case generation.
\end{comment}

To the best of our knowledge, this paper makes several novel contributions summarized as follows.
\begin{itemize}
    \item We formalize a stateful DL system as a MDP, which is able to characterize the internal states and 
    dynamic behaviors of RNN-based stateful DL systems.
    
    \item Based on the MDP abstraction, we design a set of specialized testing criteria for 
    stateful DL systems, including two state-level criteria and three transition-level criteria.
    This is the very first set of testing criteria specially designed for RNNs.

    \item We evaluate the usefulness of the criteria on a real-world ASR application, and confirm 
    that more precise abstraction can better discriminate different test sequences, and generating tests towards increasing coverage is 
    helpful for defect detection.

    \item We implement a coverage-guided testing framework, \tool.
    As the first testing framework for audio-based DL systems, we also designed a set of 
    metamorphic transformations tailored for audio inputs, inspired by real-world scenarios such as
    background noise, volume variation, etc.
    Experimental results demonstrate the effectiveness of \tool in terms of generating high-coverage tests and discovering defects on practical ASR systems.
\end{itemize}

The rest of this paper is organized as follows.
Section~\ref{sec:background} introduces the background of RNN-based stateful deep learning systems 
and one of its successful application domain, ASR, concerned in this paper.
Section~\ref{sec:overview} presents the high-level workflow of our coverage-guided testing 
framework and tool implementation \tool for stateful DL systems.
Section~\ref{sec:trans_cov} discusses the detailed state transition modeling and formalization of 
RNN and Section~\ref{sec:criteria} proposes the testing criteria specialized for RNN-based stateful 
DL systems. 
Based on these, Section~\ref{sec:framework} proposes a coverage-guided testing framework for defect 
detection in stateful DL systems.
Section~\ref{sec:evaluation} performs a large scale evaluation of our proposed technique on a 
practical end-to-end RNN-based ASR system. 
Finally, we discuss the related work in Section~\ref{sec:related} and concludes the paper with some 
possible future directions in Section~\ref{sec:conclusion}.

%% file: background.tex
\section{Background}
\label{sec:background}

\begin{figure}[t]
    \centering
    \includegraphics[width=0.95\columnwidth]{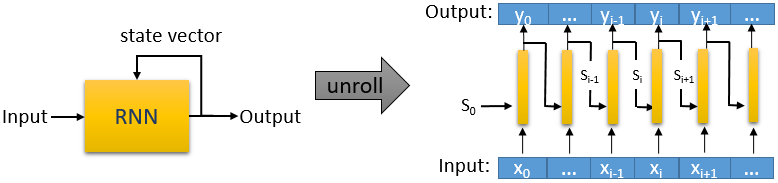}
    \caption{Architecture of simplified RNN.}
    \label{fig:rnn_arch}
\end{figure}

In this section, we introduce background of RNN and its applications in automated speech recognition.

\subsection{Recurrent Neural Network}
Different from the feed-forward DL systems (e.g, CNN), Recurrent Neural Networks (RNNs) implement 
temporal behaviors with loops and memorization with internal states to take into account influence of previous~(or future) 
observations.
Such stateful nature of RNN contributes to its huge advantage and success in handling sequential 
inputs, and leads to its domination in current industrial applications on audio, natural languages 
and video processing, making huge impact on our daily life.
A simplified version of the RNN architecture is illustrated in Figure~\ref{fig:rnn_arch}.
A basic RNN is a network of neuron-like nodes organized into successive \emph{iterations}~(or \emph{loops}).
It takes as inputs both the data stream to process and the internal state vector maintained.
Instead of taking the input data as a whole, RNN processes a small chunk of data as it arrives, and 
sequentially produces outputs at each iteration and updates the state vector.
For each individual input sequence, the state vector is first initialized to $s_0$ (usually a vector of 
zeros), and the state vector from previous iteration is passed to the next iteration.
At the high level, the sequence of state vectors can be seen as a trace recording the underlying 
temporal dynamics of RNNs, thus providing a witness for the overall characteristics of the network.

With the loop design, the ``vanishing gradient problem''~\cite{hochreiter2001gradient} becomes more 
severe on RNN, where gradient used in back propagation of training can either vanish to zero or 
becomes extremely large when the number of iterations increases, causing the model difficult to 
be optimized.
Long Short-Term Memory (LSTM)~\cite{hochreiter1997long} and Gated Recurrent Unit 
(GRU)~\cite{cho2014properties} are designed to overcome this problem.
Their network structures are much more complicated than the basic RNN shown in 
Figure~\ref{fig:rnn_arch}, but all shares the simple basic principle of RNN design and implements the state vectors for memorization.

\subsection{Automated Speech Recognition}

\begin{figure}[t]
    \centering
    \includegraphics[width=0.8\columnwidth]{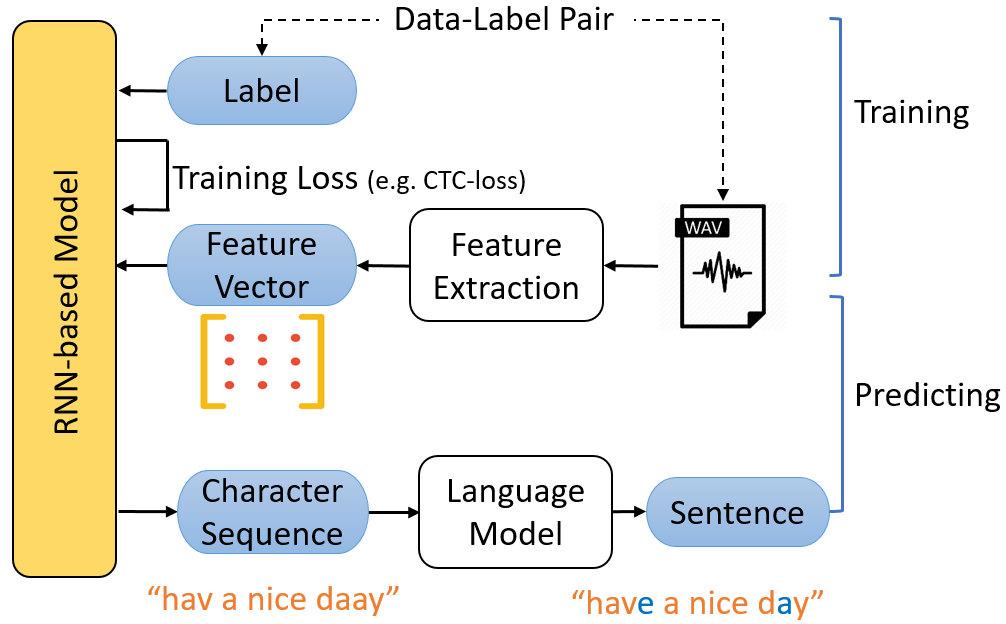}
    \caption{General training and predicting workflow of end-to-end ASR systems.}
    \label{fig:asr_arch}
\end{figure}

ASR plays an important role in intelligent voice assistants, e.g., Amazon Alexa, Google 
Assistant and Apple Siri.
The essential component of an ASR is a \emph{transcribing module} responsible for converting speech 
features to texts.
Historically, developing the transcribing module requires significant manual efforts in deriving the 
phonetic alignment of audios and texts, which severally limits the scale of the training data.
With advanced training loss functions, such as ``Connectionist Temporal Classification'' (CTC) 
loss~\cite{graves2006connectionist}, the alignment can be handled internally by an end-to-end 
approach.
The transcribing modules in current end-to-end ASRs are mostly equipped with a RNN kernel which may 
also be accompanied by some auxiliary CNN layers.
Most popular open source end-to-end ASR systems, including 
DeepSpeech~\cite{Hannun2014} and EESEN~\cite{miao2015eesen}, are powered by RNN-based models.

A general workflow of the end-to-end ASR system is demonstrated in Figure~\ref{fig:asr_arch}, 
with training and predicting procedures highlighted, respectively.
These two procedures share the same \emph{feature extraction module}, where a raw audio is 
transformed to numerical features (e.g., Mel-Frequency Cepstral Coefficients (MFCC) features).
During the training stage, feature vectors and text transcriptions are fed directly into the 
RNN-based model, skipping the acoustic model for phonetic analysis.
As for prediction, the pipeline is composed of three major components, a \emph{feature extraction 
module}, a \emph{RNN-based translation module} and a \emph{language model-based correction module}.
Usually, the \emph{language model} is provided externally and requires no training.
Audios are first transformed into numerical feature vectors, and then passed to the DNN model to 
obtain initial transcriptions, which can be further refined or corrected by the \emph{language model}.
For example, in Figure~\ref{fig:asr_arch}, the spelling errors in the character sequence are 
easily rectified.

%% file: overview.tex
\section{The Overview of Coverage-guided Testing Framework \tool}
\label{sec:overview}

\begin{figure}[t]
    \centering
    \includegraphics[width=.85\columnwidth]{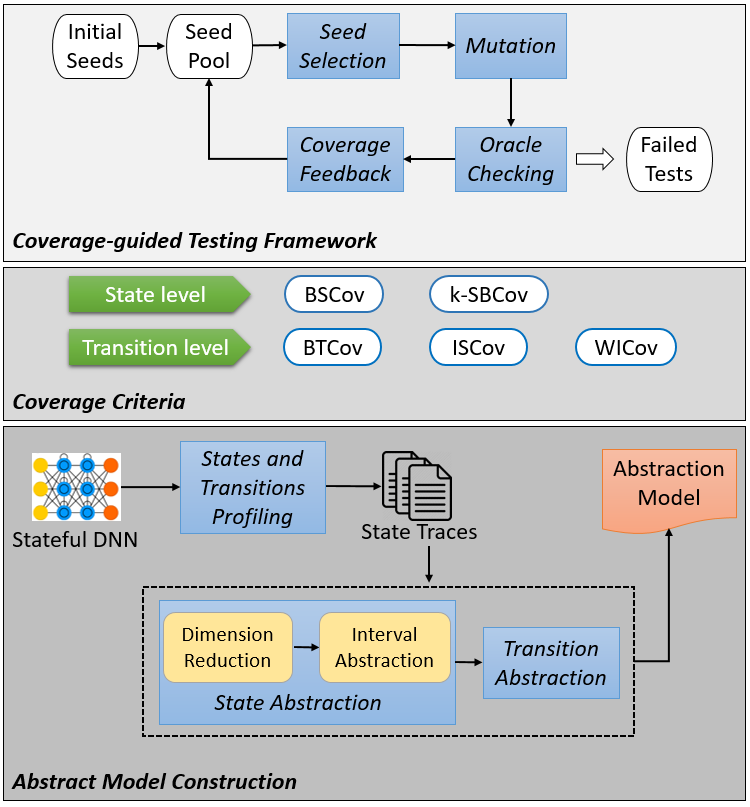}
    \caption{High-level workflow of \tool.}
    \label{fig:approach_overview}
\end{figure}

Figure~\ref{fig:approach_overview} summarizes the workflow of our approach, including the 
\emph{abstract model construction} of RNN, the \emph{coverage criteria} defined over the RNN model, 
and a \emph{coverage-guided automated testing framework} to generate tests for defect and 
vulnerability detection of RNNs.

The abstract model construction module takes a trained RNN as input and analyzes its internal 
runtime behaviors through profiling.
The quality of the profiling largely depends on the input data used.
The ideal choice of inputs for profiling is the training data (or part of them), which best reflect 
the internal dynamics of a trained RNN model.
Specifically, each input sequence is profiled to derive a \emph{trace}, i.e., a sequence of RNN 
state vectors.
After all the provided inputs are profiled, we will get a set of traces which describe the major 
states visited and transitions taken by an RNN.

In practice, the internal state space of an RNN and the number of traces enabled by the training 
data are often beyond our analysis capability.
Therefore, we perform abstraction over the states and traces to obtain an abstract model capturing 
the global characteristics of the trained network.
At the state level, we apply Principle Component Analysis (PCA)~\cite{jolliffe2011principal} to 
reduce the dimensions of the vectors and keeps the first $k$ dominant components.
For each of the $k$ dimensions, we further partition them into $m$ equal intervals.
At the transition level, we consolidate concrete transitions into abstract ones according to the 
state abstraction.
We also take into account the frequencies of different transitions under various inputs and 
effectively derive a Markov Decision Process (MDP)~\cite{Puterman:1994:MDP:528623} model for the 
trained RNN.

Based on the abstract model, five coverage criteria are then designed to facilitate the systematic testing of RNN.
These include two state-level coverage criteria -- the \emph{basic state coverage} (BSCov) and the 
\emph{$k$-step state boundary coverage} ($k$-SBCov), and three transition-level criteria -- the 
\emph{basic transition coverage} (BTCov), the \emph{input space coverage} (ISCov) and the 
\emph{weighted input coverage} (WICov).
These criteria are used to guide test case generation in our evaluation with the aim to uncover 
defects in the RNN under test (see Section~\ref{sec:evaluation}).

The testing framework is designed to make use of the above criteria to facilitate the defect discovery of RNN-based systems.
With a chosen coverage criterion, the testing process starts with a set of initial seeds (some 
audios under practical ASR testing scenarios), namely the \emph{seed pool}.
Then in each iteration, we select an audio based on heuristics and generate a mutant from 
it by applying certain transformations.
If the generated mutant triggers some defects, e.g., the transcription shows a large difference 
from the expectation,
it is labeled as a failing test.
Otherwise, we check whether the mutant improves the test coverage with respect to the chosen 
criteria and include it into the seed pool if so.

%% file: modeling.tex
\section{State Transition Modeling of Recurrent Neural Network}
\label{sec:trans_cov}

RNN models are inherently stateful~\cite{omlin1996constructing}.
In this section, we formalize the internal states and state transitions of RNNs, and describe an 
abstract model used to capture the global characteristics of the trained RNN models.

\subsection{RNN Internal States and State Transitions}
Following \cite{rastogi2016weighting}, we represent a neural network abstractly as a differentiable 
parameterized function $f(\cdot)$.
The input to a RNN is a sequence $\vec{x} \in \inputs^N$, where  \inputs is the input domain and 
$N$ is the length of the sequence.
Let $x_i \in \inputs$ be the $i$-th element of that sequence.
Then, when passing $\vec{x}$ into a RNN, it maintains a state vector $\vec{s} \in \states^N$ with 
$s_0 = \vec{0}$ and $(s_{i+1}, y_i) = f(s_i, x_i)$, where \states is the domain of the hidden 
state, $s_i \in \states$ is the hidden state of RNN at the $i$-th iteration, and $y_i \in \outputs$ 
is the corresponding output at that step.
We use $s_i^d$ to denote the $d$-th dimension of the state vector $s_i$.

Naturally, each input sequence $\vec{x}$ induces a finite sequence of state transitions $\vec{t}$, 
which we define as a \emph{trace}.
The $i$-th element in a trace $\vec{t}$, denoted by $t_i$, is the transition from $s_i$ to 
$s_{i+1}$ after accepting an input $x_i$ and producing an output $y_i$.
A Finite State Transducer (FST)~\cite{gill1962introduction} can be used to represent a collection 
of traces more compactly~\cite{Horne98finitestate} as defined below.

\begin{definition}
A FST is a tuple $(\states, \inputs, \outputs, I, F, \trans)$ such that \states is a 
non-empty finite set of states, \inputs is the input alphabet, \outputs is the output alphabet, $I 
\subseteq \states$ is the set of initial states, $F \subseteq \states$ is the set of final states, 
and $\trans \subseteq \states \times \inputs \times \outputs \times \states$ is the transition 
relation.
\end{definition}

\begin{figure}
	\centering
	\includegraphics[width=.65\columnwidth]{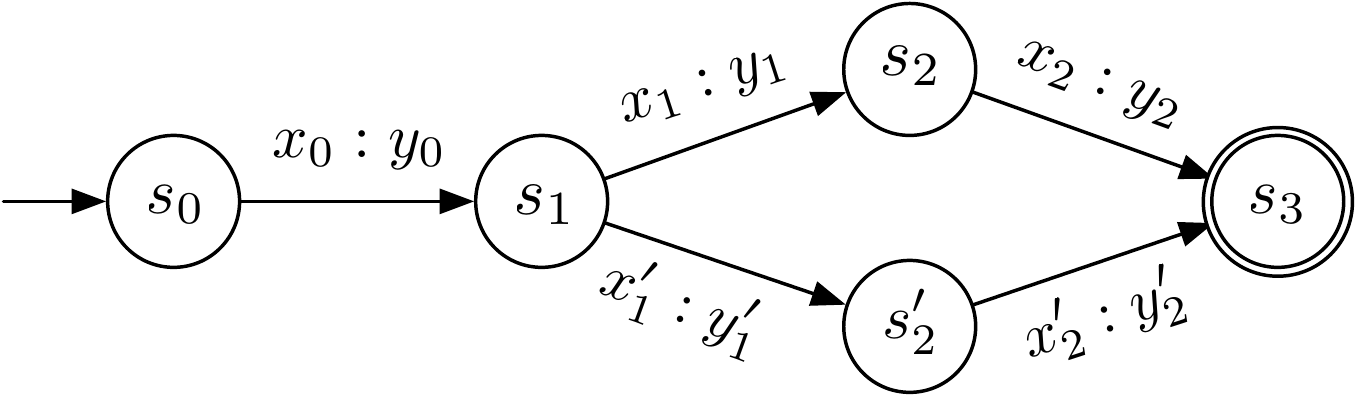}
	\caption{An example FST representing two traces.}\label{fig:fst-exp}
\end{figure}

For example, Fig.~\ref{fig:fst-exp} shows a simple FST representing two traces, namely,
$s_0s_1s_2s_3$ and $s_0s_1s_2's_3$ with $s_0$ being the initial state and $s_3$ being the 
final state.
The first trace takes an input sequence $x_0x_1x_2$ and emits an output sequence $y_0y_1y_2$; the 
second trace takes an input sequence $x_0x_1'x_2$ and emits an output sequence $y_0y_1'y_2$.

\subsection{Abstract State Transition Model}
The number of states and traces enabled while training a RNN can be huge.
To effectively capture the behaviors triggered by a large number of input sequences and better 
capture the global characteristics of the trained network, we introduce an \emph{abstract state 
transition model} in this paper.
The abstract model over-approximates the observed traces induced of an RNN and has a much 
smaller set of states and transitions compared with the original one.
The abstraction is also configurable -- one can trade-off between the size and precision of the 
model so that the abstract model is still able to maintain useful information of the input 
sequences for particular analysis tasks.
To obtain an abstract model for a trained RNN, we abstract over both the states and the transitions.

\paragraph{State Abstraction}
Each \emph{concrete state} $s_i$ is represented as a vector $(s_i^1, \ldots, s_i^m)$, usually in 
high dimension (i.e., $m$ could be a large number).
Intuitively, an \emph{abstract state} represents a set of concrete states which are close in space.
To obtain such a state abstraction, we first apply the Principle Component Analysis 
(PCA)~\cite{jolliffe2011principal} to perform an orthogonal transformation  on the concrete 
states -- finding the first $k$ principle components~(i.e., axes) which best distinguish the given state vectors and ignore 
their differences on the other components.
This is effectively to project all concrete states onto the chosen $k$-dimensional component basis.

Then, we split the new $k$-dimensional space into $m^k$ \emph{regular 
grids}~\cite{thompson1998handbook} such that there are $m$ equal-length intervals on each axis: 
where $e_i^d$ represents the $i$-th interval on the $d$-th dimension, $lb_d$ and $ub_d$ are the 
lower and upper bounds of all state vectors on the $d$-th dimension, respectively.
In this way, all concrete states $s_i$ which fall within the same grid are mapped to the same abstract state: 
$
\abss = \{s_i | s_i^1 \in e_{\_}^1 \wedge \cdots \wedge s_i^k \in e_{\_}^k \}.
$
We denote the set of all abstract states as \astates.
Noticeably, the precision of the state abstraction can easily be configured by tuning the 
parameters $k$ and $m$.

Let $j = I^d(\abss)$ be the index of \abss on the $d$-th dimension such that for all 
$s \in \abss$, $s^d$ falls in $e_j^d$ ($0 \leq j < m$).
For any two abstract states \abss and $\abss'$, we define their \emph{distance} as:
$$
Dist(\abss, \abss') = \Sigma_{d=1}^k |I^d(\abss) - I^d(\abss')|.
$$
This definition can also be generalized to include space beyond the lower and upper bounds.

\begin{figure}
	\centering
	\begin{subfigure}{.55\columnwidth}
		\includegraphics[width=\linewidth]{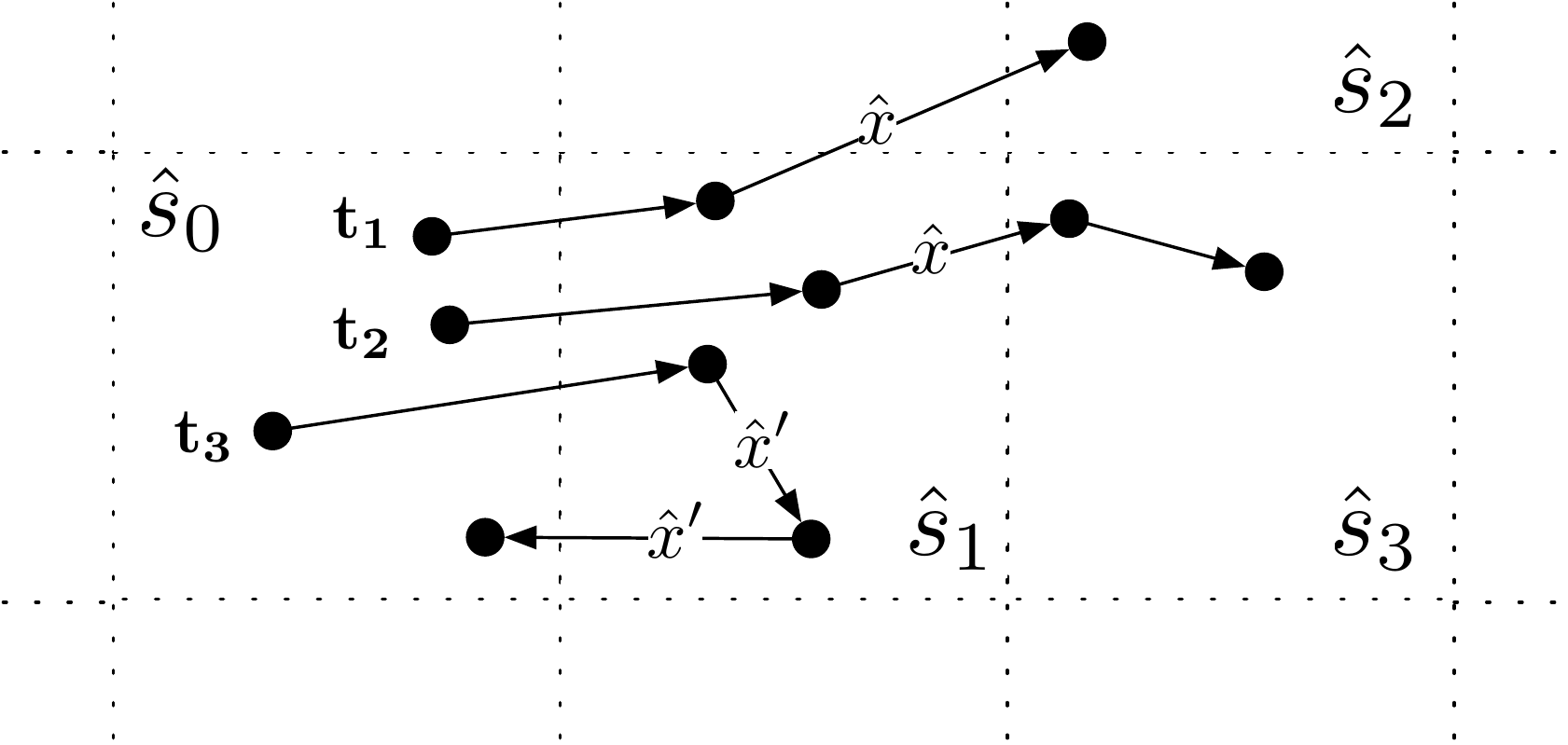}
		\caption{Concrete traces.}\label{fig:con}
	\end{subfigure}
	\begin{subfigure}{.45\columnwidth}
		\includegraphics[width=\linewidth]{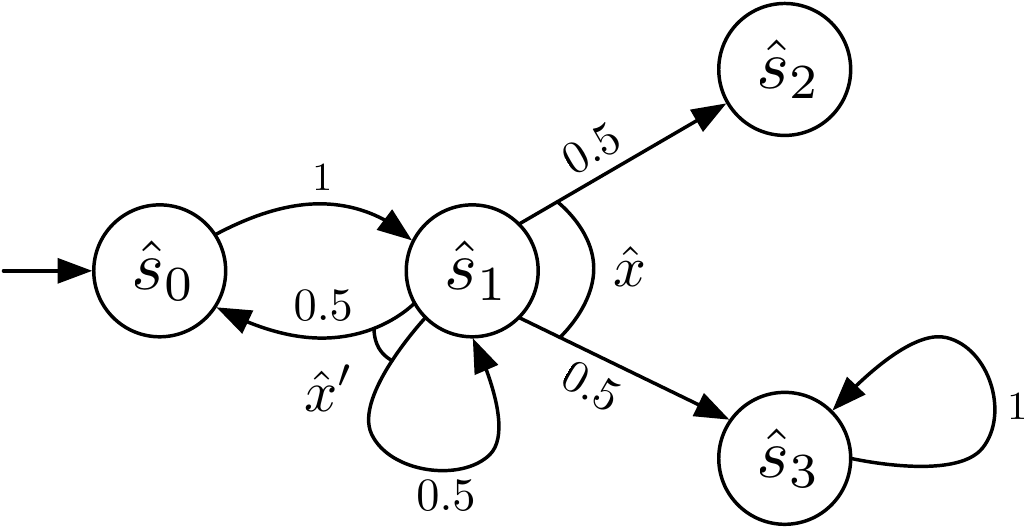}
		\caption{MDP abstraction.}\label{fig:mdp}
	\end{subfigure}
	\caption{A set of concrete traces and their corresponding abstract state transition 
		model.}\label{fig:mdp-exp}
\end{figure}

\paragraph{Transition Abstraction}
Once the state abstraction is computed, a concrete transition between two concrete states can be 
mapped as a part of an \emph{abstract transition}.
An abstract transition represents a set of concrete transitions which share the same source and 
destination abstract states.
In other words, there is an abstract transition between two abstract states $\abss$ and $\abss'$ if
and only if there exists a concrete transition between $s$ and $s'$ such that $s \in \abss \wedge 
s' \in \abss'$.
The set of all abstract transitions is denoted as $\atrans \subseteq \astates \times \astates$.

For instance, Figure~\ref{fig:con} depicts three concrete traces, i.e., $\vec{t}_1$, $\vec{t}_2$ 
and $\vec{t}_3$, where states are shown as dots and transitions are directed edges connecting dots.
The grids drawn in dashed lines represent the abstract states, i.e., $\abss_0$, $\abss_1$, 
$\abss_2$, and $\abss_3$, each of which is mapped to a set of concrete states inside the corresponding grid.
The set of abstract transitions is, therefore, $\{(\abss_0,\abss_1), (\abss_1,\abss_0), 
(\abss_1,\abss_1),(\abss_1,\abss_2),(\abss_1,\abss_3),(\abss_3,\abss_3)\}$.

\subsection{Representing Trained RNN as a Markov Decision Process}
Each input sequence in the training set yields a concrete trace of the RNN model.
The abstract state transition model captures all the concrete traces enabled from training data~(or its representative parts) and other potential traces which have not been enabled.
Because of the ways how the state and transition abstractions are defined, the resulting abstract 
model represents an over-approximation and generalization of the observed behaviors of the trained RNN model.

To also take into account the likelihood of abstract transitions under different inputs, we augment the abstract model with transition probabilities and non-deterministic choices, effectively making it a Markov Decision Process (MDP)~\cite{Puterman:1994:MDP:528623} without costs.

\begin{definition}
A MDP is a tuple $(\astates, I, \atransfun)$, where \astates is a set of abstract states, $I$ is a 
set of initial states, and $\atransfun: \astates \times \ainputs \mapsto Dist(\astates)$ is the 
\emph{transition probability function} such that \ainputs represent the (abstract) input space and 
$Dist(\astates)$ is the set of discrete probability distributions over the set of abstract states.
\end{definition}

The space of choices at a state \abss is given by the set of inputs $\ainputs(\abss)$ enabled at 
that state, which is abstracted in the same way as the states.
We write $\Pr_{\absx}(\abss,\abss')$ to denote the conditional probability of visiting $\abss'$ 
given the current state \abss with input \absx, such that $\Sigma_{\abss' \in \astates} 
\Pr_{\absx}(\abss,\abss') = 1$.
We define the transition probability as the number of concrete transitions from \abss to $\abss'$ 
over the number of all outgoing concrete transitions from \abss given the input \absx, i.e.,
$$
\Pr_{\absx}(\abss,\abss') = \frac{| \{(s,s',x) |  s \in \abss \wedge x \in \absx \wedge s' \in 
	\abss'\}|}{|\{(s,\_,x) | s \in \abss \wedge x \in \absx\}|}.
$$

For example, Figure~\ref{fig:mdp} shows the abstract state transition model for the concrete traces 
in Figure~\ref{fig:con} as a MDP.
The abstract transitions are labeled with their transition probabilities.
For instance, since all outgoing transitions at $\abss_0$ end in $\abss_1$, the transition 
probability from $\abss_0$ to $\abss_1$ is one.
There are two choices of inputs at $\abss_1$, i.e., $\ainputs(\abss_1) = \{\absx,\absx'\}$.
When the given input at $\abss_1$ is \absx, the transition probabilities are computed as 
$\Pr_{\absx}(\abss_1,\abss_2) = \frac{1}{2}$ and $\Pr_{\absx}(\abss_1,\abss_3) = \frac{1}{2}$.
The computation for the case when the input is $\absx'$ is similar.

As is shown in the example, a MDP model is constructed by first applying the state and transition 
abstractions on a set of concrete traces, and then computing transition probability distributions 
for each input at every abstract state.
The time complexity of the abstraction step depends on the number of concrete traces, while the 
complexity for computing the transition probabilities only depends on the number of abstract 
transitions.

%% file: criteria.tex
\section{Coverage Criteria of Stateful Recurrent Neural Network}\label{sec:criteria}
Inspired by traditional software testing, we propose a set of testing coverage criteria for RNNs 
based on the abstract state transition model.
The goal of the RNN coverage criteria is to measure the completeness and thoroughness of test data 
in exercising the trained as well as the unseen behaviors.
The state and transition abstractions are designed to reflect the internal network configurations 
at a certain point as well as the temporal behaviors of the network over time, respectively.
Therefore, to maximize the chance of discovering defects in stateful neural networks, one should 
combine coverage criteria based on both the state and transition abstractions to systematically 
generate comprehensive and diverse test suites.

Let $M = (\astates, I, \atransfun)$ be an abstract model of the trained RNN represented as a MDP.
Let $T = \{\vec{x}_0, \ldots, \vec{x}_n\}$ be a set of test input sequences.
We define both the \emph{state-level} and \emph{transition-level} coverage of $T$ to measure how 
extensively $T$ exercises the states and transitions of $M$, respectively.

\subsection{State-Level Coverage Criteria}
The state-level coverage criteria focuses on the internal states of the RNN.
The set of abstract states \astates represents a space generalization of the visited states obtained from training data~(or its representative parts), which is referred to as the \emph{major function region}~\cite{ma2018deepgauge}.
The space outside the major function region is never visited by the training data, and thus 
represents the \emph{corner-case region}~\cite{ma2018deepgauge}.
The test data should cover the major function region extensively to validate the trained behaviors and cover the corner-case region sufficiently in order to discover defects in unseen behaviors.

\paragraph{Basic State Coverage}
Given a RNN abstract model $M$ and a set of test inputs $T$, the \emph{basic state coverage} 
measures how thoroughly $T$ covers the major function region visited while training.
To quantify this, we compare the set of abstract states visited by the training inputs and the test 
inputs, denoted by $\astates_{M}$ and $\astates_{T}$, respectively.
Then the basic state coverage is given by the number of abstract states visited by both the 
training and the test inputs over the number of states visited by the training inputs:
$$
\textsc{BSCov}(T,M) = \frac{|\astates_{T} \cap \astates_{M}|}{|\astates_{M}|}.
$$

\paragraph{$k$-Step State Boundary Coverage}
The test data may also trigger new states which are never visited during training.
The \emph{$k$-step state boundary coverage} measures how well the corner-case regions are covered 
by the test inputs $T$.
The corner-case regions $\astates_{M^c}$ are the set of abstract states outside of $\astates_{M}$, 
which have non-zero distances from any states in $\astates_{M}$.
Then $\astates_{M^c}$ can be further divided into different boundary regions defined by their 
distances from $\astates_{M^c}$.
For example, the \emph{$k$-step boundary region}, $\astates_{M^c}(k)$, contains all abstract states 
which have a minimal distance $k$ from $\astates_{M}$, or more formally,
$
\astates_{M^c}(k) = \{\abss \in \astates_{M^c} | \min_{\abss' \in \astates_{M}} Dist(\abss, \abss') 
= k \}.
$

The $k$-step state boundary coverage is defined as the ratio of states visited by the test inputs 
in the boundary regions of at most $k$ steps away from $\astates_{M}$:
$$
k\text{-}\textsc{SBCov}(T,M) = \frac{|\astates_{T} \cap \bigcup_{i=1}^k \astates_{M^c}(i)|}%
{|\bigcup_{i=1}^k \astates_{M^c}(i)|}.
$$

\subsection{Transition-Level Coverage Criteria}
The state-level coverage indicates how thorough the internal states of an RNN are exercised but it 
does not reflect the different ways transitions have happened among states in successive time steps.
The transition-level coverage criteria targets at the abstract transitions activated by various 
input sequences and a higher transition coverage shows that the inputs are more adequate in 
triggering diverse temporal dynamic behaviors.

\paragraph{Basic Transition Coverage}
To quantify transition coverage, we compare the abstract transitions exercised during both the 
training and testing stages, written as $\atrans_{M}$ and $\atrans_{T}$, respectively.
Then the \emph{basic transition coverage} is given by:
$$
\textsc{BTCov}(T,M) = \frac{|\atrans_{T} \cap \atrans_{M}|}{|\atrans_{M}|}.
$$
The basic transition coverage subsumes the basic state coverage.
In other words, for any abstract model $M$, every test input $T$ satisfies basic transition 
coverage with respect to $M$, also satisfies the basic state coverage.

\paragraph{Input Space Coverage}
The input space at an abstract state \abss is given as $\ainputs_{\abss}$, which represents the set 
of abstract inputs accepted at \abss while training.
The test inputs should cover the input space at each state as much as possible to exercise the 
different subsequent transitions.
More formally, let the input spaces for the training and test data at \abss be $\ainputs_M(\abss)$ 
and $\ainputs_T(\abss)$, respectively. 
Then the \emph{input space coverage} is defined as:
$$
\textsc{ISCov}(T,M) = \frac{\Sigma_{\abss \in \astates_{T} \cap \astates_{M}} 
|\ainputs_T(\abss)|}{\Sigma_{\abss \in \astates_{M}} |\ainputs_M(\abss)|}.
$$
Note that the input space coverage is incomparable to the basic transition coverage -- achieving 
the input coverage does not guarantee the transition coverage, and vice versa.

\paragraph{Weighted Input Coverage}
Our abstract model also encodes the frequencies of different transitions given a particular input, 
observed during training.
More frequently triggered transitions have a higher transition probability, which is given by the 
transition probability function \atransfun.
The \emph{weighted input coverage} considers not only the different choices at a state, but also 
the range of possible subsequent transitions when a specific input is chosen.
More formally, it is defined as:
$$
    \textsc{WICov}(T,M) = \frac{
		  \Sigma_{\abss \in \astates_{T} \cap \astates_{M}} 
		  \Sigma_{\absx \in \ainputs_T(\abss) \cap \ainputs_M(\abss)}
		  \Sigma_{\abss' \in \atrans_{T}(\abss)} \atransfun(\abss,\absx,\abss')
	  }{\Sigma_{\abss \in \astates_{M}} |\ainputs_M(\abss)|}
$$
The weighted input coverage is stronger than both the basic transition coverage and the input space 
coverage.

%% file: testing_framework.tex
\section{Coverage-Guided Testing Framework}\label{sec:framework}
In this section, we introduce the coverage-guided testing framework.
We first describe a group of metamorphic transformations specialized for audio signals (Section~\ref{sub:mras}) and then present a mutation-based test generation algorithm which is guided by 
the coverage criteria proposed in Section~\ref{sec:criteria}.

\subsection{Metamorphic Transformations of Audio Signals}\label{sub:mras}
\label{sec:mutation}
ASR performs general transformation from audio speeches to the corresponding natural-language texts, and is often expected to work properly on speeches with various volume, speed and voice characteristics.
Also, these speech audio signals may be mixed with noises coming from ambient sounds and not well-insulated receivers.
Inspired by these practical scenarios, we derive a set of transformation operators to mimic the environment interference.
Overall, they can be classified into four categories:
\begin{itemize}[leftmargin=*]
\item{\it Volume-related transformations (VRT):} ChangeVolume, LowPassFilter, HighPassFilter.
\item{\it Speed-related transformations (SRT):} PitchShift, ChangeSpeed.
\item{\it Clearness-related transformations (CRT):} AddWhiteNoise, ExtractHarmonic
\item{\it Unaffected transformations (UAT):} DRC, Trim.
\end{itemize}
Categories {\it VRT, SRT} and {\it CRT} affect the volume, speed and clearness of an audio signal, respectively; 
and Category {\it UAT} contains transformations that affect neither of them, but still makes minor changes to the speech signal.
Table~\ref{tb:transformation} summarizes the transformations with brief descriptions.

For defect detection, our goal is to generate audios which sounds normal to human but are incorrectly transcribed by ASRs. 
With a diverse collection of transformations, an audio can be mutated to generate new audios,
among which there could be ones trigger new traces in the trained network and lead to potential defects in the ASR.
However, a violent transformations with significant perturbations may result in an audio which is not recognizable even by human. 
For instance, the volume may become too low or the frequency may become too high.
To generate suitable defect-triggering candidates, we apply the transformations with restraints to ensure that the audio seeds and the corresponding mutants sound the same to human beings.
Transformation operators satisfying the above requirements are said to keep a \emph{metamorphic relation}~\cite{chen1998metamorphic} and we refer to them as \emph{metamorphic transformations}.
Now we propose a strategy to perform transformations made by \tool with the best effort to preserve its text information of the audio before and after transformation.

\begin{table}[t]
\centering
\caption{Transformations for audio signals.}
\label{tb:transformation}
\begin{tabular}{lll}
\toprule
   & Transformation             & \multicolumn{1}{c}{Description}                                                                                                                                                                                 \\ \hline
1  & \textit{AddWhiteNoise}     & Randomly add white noise in the audio                                                                                                                                                                            \\
2  & \textit{PitchShift}        & \begin{tabular}[c]{@{}l@{}}Pitch-shift the waveform of the audio to \\ raise or lower the pitch of an audio signal\\  by a random interval\end{tabular}                                                          \\ 
3  & \textit{Trim}              & \begin{tabular}[c]{@{}l@{}}Trim leading and trailing silence from\\  an audio signal\end{tabular}                                                                                                                \\ 
4  & \textit{ChangeSpeed}       & Randomly speed up or slow down the audio                                                                                                                                                                         \\ 
5  & \textit{ChangeVolume}      & Randomly adjust the volume of the audio                                                                                                                                                                          \\ 
6  & \textit{DRC}               & \begin{tabular}[c]{@{}l@{}}Dynamic range compression (DRC), that \\ reduces the volume of loud sounds or \\ amplifies quiet  sounds thus reducing or \\ compressing an audio signal's dynamic range\end{tabular} \\ 
7  & \textit{LowPassFilter}     & \begin{tabular}[c]{@{}l@{}}Pass signals with a frequency lower than a \\ random selected cutoff frequency and \\ attenuates signals with frequencies higher \\ than the cutoff frequency\end{tabular}            \\ 
8 & \textit{HighPassFilter}    & \begin{tabular}[c]{@{}l@{}}Pass signals with a frequency higher than a \\ random cutoff frequency and attenuates \\ signals with frequencies lower than the \\ cutoff frequency\end{tabular}                     \\ \bottomrule
\end{tabular}
\end{table}

In general, volume, speed and clearness transformations have no effect on the audio semantics, when applied individually and controlled under a certain degree.
However, when an audio input $a_0$ is processed with a sequence of transformations (e.g., $a_0\hookrightarrow a_1,\ldots,\hookrightarrow a_i$),
there is a higher chance that the audio cannot be well recognized by human due to accumulated distortions.
We propose a strategy to conservatively select transformations for generating defect candidates indistinguishable to human before and after the transformations:
(1) we carefully set the parameters to ensure that a single step transformation (on volume, speed or clearness) does not violate the metamorphic relations (i.e., human hearing are not affected); and (2) an audio, generated by mutating from a seed after a sequence of transformations $T$, is limited to be mutated by a transformation $t \in S \cup \mathit{UAT}$ such that $ (\{t\} \cup S) \cap T=\emptyset$, where $S\in \{VR, \mathit{SRT}, \mathit{CRT}\}$.
Intuitively, the constraints make sure that a mutant is generated by altering the volume, speed or clearness of a seed input at most once.
If this constraints is unsatisfied, the audio will not be transformed further.
%Thus, the conservative strategy leverages the metamorphic relations with the best effort preserve the execution results~(i.e., from human's perspective) for inputs before and after transformation.

\subsection{Coverage-Guided Testing}
\label{subsec:cg_testing}

Algorithm~\ref{algo1} presents the general procedure to test RNN-based DL systems with various configurable feedback.
It is consistent with the testing framework diagram shown in Fig.~\ref{fig:approach_overview}.
The inputs of \tool include initial seeds $I$ and the RNN-based deep learning system $R$. 
The outputs are regression tests with higher coverage and failed tests that are incorrectly handled. 
The initial test queue only contains initial seeds. 
For each time, \tool selects one input $a$ (i.e., an speech audio) from the test queue. 
Based on the already selected transformations of $a$, \tool randomly picks one transformation $t$ (c.f. Section~\ref{sec:mutation}). 
If no further transformation is allowed to be picked (i.e., $t = null$) under the metamorphic relation constraints, \tool will select next input from the queue. 
For a picked transformation $t$, a random parameter is picked. 
\tool will generate a new audio $a'$ with transformation $t$ and perform the transformation with the deep learning system $R$. 
If the prediction result is inconsistent with the original seed, $a'$ will be added into the failed tests. For audio in ASR, we decide a failed test by checking whether the Word/Character Error Rate (W/CER) exceeds a certain threshold. 
If the mutated test is not a failed test and increases the overall coverage, \tool adds it into the test queue and updates the coverage of tests in the test queue.

\begin{algorithm}[t]

	\footnotesize
	\SetKwInOut{Input}{input}\SetKwInOut{Output}{output}
	\SetKwInOut{Const}{const}
	\SetKwInOut{Continue}{continue}
	\Input{$I$: Initial seeds, $R$, RNN-based stateful deep learning system}
	\Output{$F$: Failed tests, $Q$: Test queue}

	$F \leftarrow \emptyset$\;
	$Q \leftarrow I$\;
	\While{$a \leftarrow Select(Q)$}{ \label{algo1:select}
	    $t \leftarrow PickTransform(a)$ \;\label{algo1:sample}
	    \If{$t$ is null}{
	       \Continue\;
	    }
	    Randomly pick parameter $p$ for $t$ \;\label{algo1:getp}
	    $a'$ = mutate($t$, $a$)vt\;\label{algo1:mutate}
	    $cov, result \leftarrow Predicate(R, a')$\;\label{algo1:predicate}
	    \If{Failed(a', result)}{ 
	        $F\leftarrow F\bigcup\{a'\}$ \label{algo1:addfailed}
	    }\ElseIf{CoverageIncrease(cov, Q)}{
	        $Q\leftarrow Q\bigcup{a'}$\;  \label{algo1:addqueue}
	        $UpdateCoverage(Q)$\;\label{algo1:updatecov}
	    }
	}
	\caption{\tool}\label{algo1}
\end{algorithm}

%% file: evaluation.tex
\section{Evaluation}
\label{sec:evaluation}

In this section, we evaluate the effectiveness of the proposed abstract state transition model and test coverage criteria, and the performance of our coverage-guided testing framework.
Through various experiments, we aim to answer the following research questions:

\begin{table}[t]
\centering
\caption{Configurations of abstract model.}
\label{table:configure}
\begin{tabular}{lrrrr}
\toprule

Name                    & $M_{\{3,10\}}$ & $M_{\{3,20\}}$ & $M_{\{3,50\}}$ & $M_{\{3,100\}}$ \\ \midrule
\#Dimensions ($k$)      & 3        & 3        & 3        & 3        \\
\#Partitions ($m$)      & 10       & 20       & 50       & 100      \\ \midrule
\#States ($|\astates|$) & 553      & 3,291    & 40,589   & 263,106  \\ \bottomrule
\end{tabular}
\end{table}

\noindent\textbf{RQ1:} Could the abstract model distinguish internal behaviours of RNN when handling different inputs? How precise is the distinction with different abstraction configurations?

\noindent\textbf{RQ2:}
Is there a correlation between the proposed criteria and erroneous behaviors of the RNN?

\noindent\textbf{RQ3:} How effective is \tool for generating high-coverage tests?

\noindent\textbf{RQ4:} How useful is \tool for defect detection in RNN-based ASR systems?

\subsection{Datasets and Experiment Setup}
\paragraph{Datasets}
We selected Mozilla's implementation of \texttt{DeepSpeech-0.3.0}~\cite{mozilladeepspeech} which produces one of 
the state-of-the-art open source ASR models.
All our experiments were conducted on a pre-trained English model released along with 
\texttt{DeepSpeech-0.3.0}, which was trained with Fisher~\cite{cieri2004fisher}, LibriSpeech~\cite{panayotov2015librispeech}, Switchboard~\cite{godfrey1992switchboard}, and a pre-release 
snapshot of the English Common Voice training corpus~\cite{commomvoice}.
It achieves an 11\% WER on the LibriSpeech clean test corpus.
In their implementation, the RNN-based model (see the part highlighted in Fig~\ref{fig:asr_arch}) is specialized with a LSTM kernel. Its state vector has 2048 dimensional 64-bit floating point, which is the major target for abstract model construction.

\paragraph{Experiment Setup}
To construct the abstract model, we randomly selected 20\% of the training data to perform the profiling considering our limited computation resource.
Even though, after profiling, we still get a huge set of state vectors (about 90 billion 64-bit floats).
In the next step, we perform Principle Component Analysis to analyze the principle components of the vector space, based on which the abstract model and transition space are constructed.

The model abstraction parameters $k$ and $m$ for the can be configured to generate abstract models with different precision.
To analyze the potential influence on these parameters on model precision, we select four different configurations listed in 
Table~\ref{table:configure}. We use $M_{\{k, m\}}$ to represent the configuration with $k$ dimensions and $m$ partitions. The last row shows the number of states under different configurations.

\begin{figure*}[t]
    \centering
    
    \begin{subfigure}[b]{0.23\textwidth}
    \centering
    \begin{minipage}{\linewidth}
    \includegraphics[width=\textwidth]{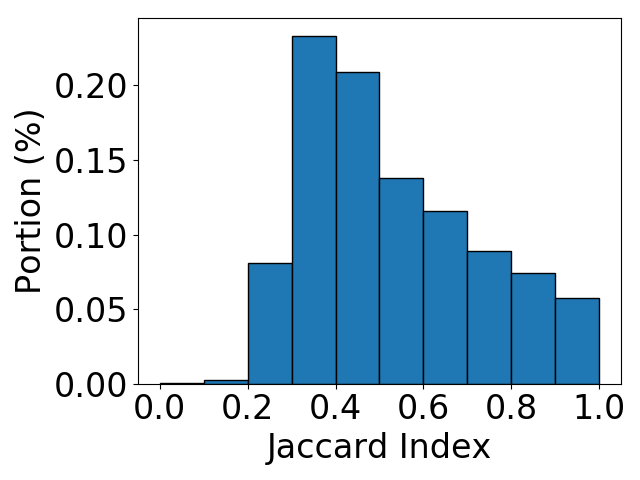}
    \end{minipage}
    \caption{Under $M_{\{3, 10\}}$}
    \end{subfigure}
    \begin{subfigure}[b]{0.23\textwidth}
    \centering
    \begin{minipage}{\linewidth}
    \includegraphics[width=\textwidth]{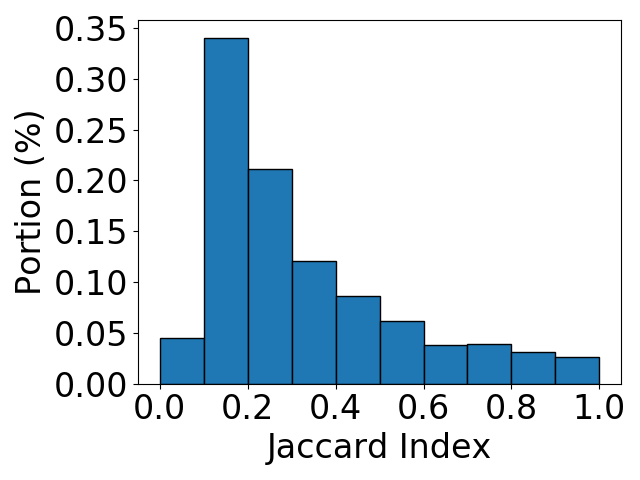}
    \end{minipage}
    \caption{Under $M_{\{3, 20\}}$}
    \end{subfigure}
    \begin{subfigure}[b]{0.23\textwidth}
    \centering
    \begin{minipage}{\linewidth}
    \includegraphics[width=\textwidth]{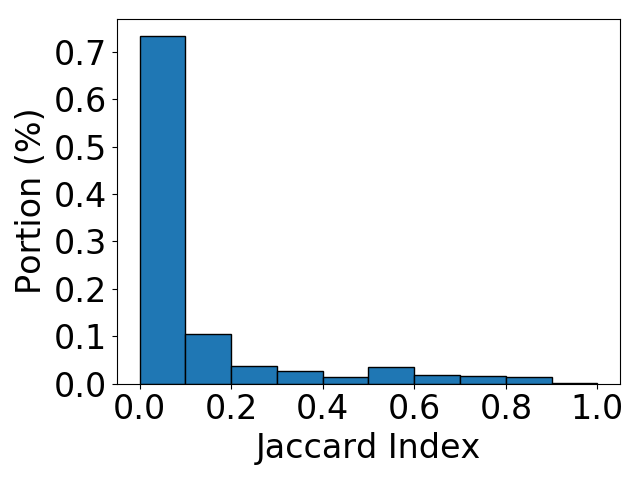}
    \end{minipage}
    \caption{Under $M_{\{3, 50\}}$}
    \end{subfigure}
    \begin{subfigure}[b]{0.23\textwidth}
    \centering
    \begin{minipage}{\linewidth}
    \includegraphics[width=\textwidth]{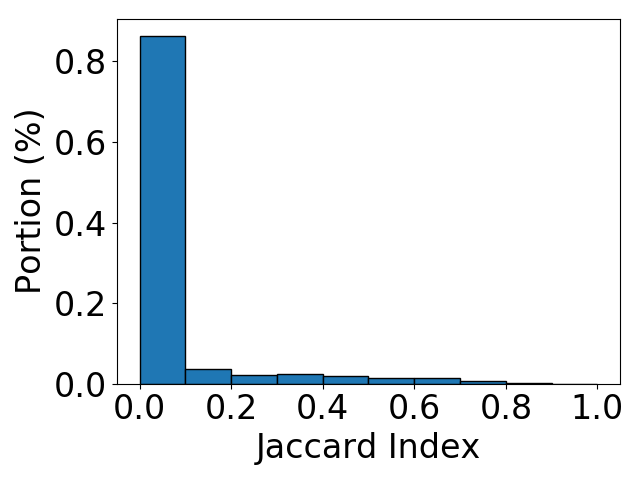}
    \end{minipage}
    \caption{Under $M_{\{3, 100\}}$}
    \end{subfigure}
    
    \caption{Results of input similarities under different abstraction configurations.}
    \label{fig:sensitivity_hist}
\end{figure*}

\begin{figure*}[t]
    \centering
    
    \begin{subfigure}[b]{0.23\textwidth}
    \centering
    \begin{minipage}{\linewidth}
    \includegraphics[width=\textwidth]{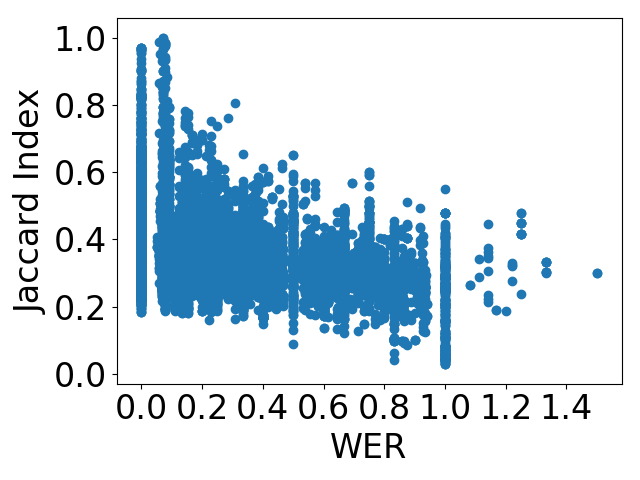}
    \end{minipage}
    \caption{Under $M_{\{3,10\}}$}
    \end{subfigure}
    \begin{subfigure}[b]{0.23\textwidth}
    \centering
    \begin{minipage}{\linewidth}
    \includegraphics[width=\textwidth]{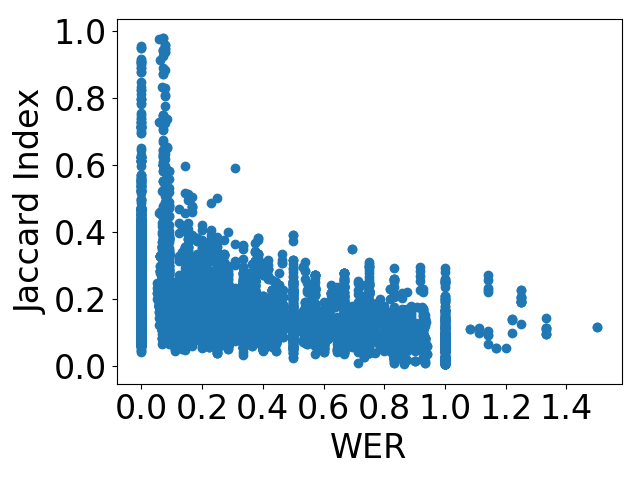}
    \end{minipage}
    \caption{Under $M_{\{3, 20\}}$}
    \end{subfigure}
    \begin{subfigure}[b]{0.23\textwidth}
    \centering
    \begin{minipage}{\linewidth}
    \includegraphics[width=\textwidth]{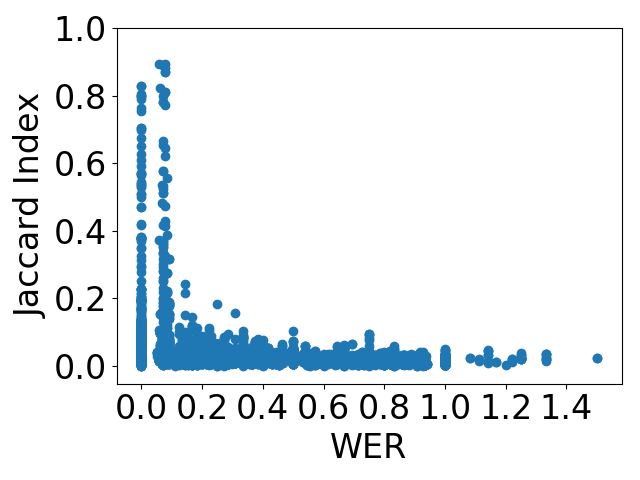}
    \end{minipage}
    \caption{Under $M_{\{3, 50\}}$}
    \end{subfigure}
    \begin{subfigure}[b]{0.23\textwidth}
    \centering
    \begin{minipage}{\linewidth}
    \includegraphics[width=\textwidth]{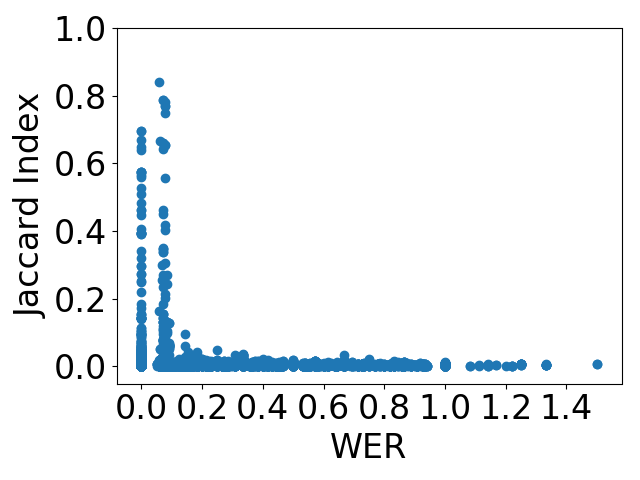}
    \end{minipage}
    \caption{Under $M_{\{3, 100\}}$}
    \end{subfigure}
    
    \caption{Correlation between basic state variation and relative WER under different abstraction configurations.}
    \label{fig:correlation}
\end{figure*}

\subsection{RQ1: Evaluation on Abstract Model Precision}
\label{subsec:rq1}
Since our testing criteria are designed based on the abstract model, the precision of the abstract model would directly influence the precision of the testing criteria in distinguishing the internal behaviours of RNN given different inputs.
An accurate parameter configuration of testing criteria would allow to distinguish on the internal behaviours of RNN on even highly similar inputs. 
To generate new tests, we perform metamorphic transformations (Section~\ref{sec:mutation}) with very small changes to keep the metamorphic constraints with the best effort. Specifically, in this evaluation, we randomly select 100 inputs from the test data. For each audio, 100 inputs are generated using different metamorphic transformations. Finally, we have 100 original input and 10,000 new inputs.

To quantify the similarity between two inputs, we adopt the Jaccard index to measure their coverage similarity. Given an abstract model $M$ and an instance $x$, we denote the set of basic states covered by $x$ as $\astates_{x}$. Then the Jaccard Index $\textsc{J}_\textsc{M}(x, y)$ for inputs $x$ and $y$ is calculated as:
$$
\textsc{J}_\textsc{M}(x, y) = \frac{|\astates_{x} \cap \astates_{y}|}{|\astates_{x} \cup \astates_{y}|}
$$
Jaccard Index ranges over $[0, 1]$, and $0$ indicates no overlapping (i.e., $x$ and $y$ is totally different) between two sets while $1$ for totally duplicate sets (i.e., $x$ and $y$ is very similar). Note that when Jaccard Index is $1$, it does not imply that $x$ and $y$ are absolutely the same as the transitions may be different. 

For each one of the new inputs and the corresponding original input, we compute a Jaccard Index value. Figure~\ref{fig:sensitivity_hist} shows the distribution of the 10,000 Jaccard Index values under different abstraction configurations. Under the configuration $M_{\{3, 10\}}$, some of inputs cannot be be well differentiated as the configuration is too coarse. For exmaple, more than 90\% of Jaccard Index values are greater or equal than 0.3, and even more than 5\% values are distributed in the range $[0.9, 1.0]$. After the abstraction is refined (i.e., the grids are more fine-grained), most of the Jaccard Index values become smaller. For example, under $M_{\{3, 100\}}$, more than 85\% of values are distributed in the range $[0.0, 0.1]$.   

\begin{tcolorbox}[size=title]
{\textbf{Answer to RQ 1:}} The abstract model can distinguish internal behaviors of RNN for different inputs effectively, even for the inputs with small differences (e.g., the inputs with metamorphic relations). The accuracy of abstract models under different abstraction configurations also varies. Abstraction with more fine-grained grids distinguishes the outputs better.
\end{tcolorbox}

\subsection{RQ2: Correlation Between Testing Criteria and Erroneous Behaviors of the RNN}
With RQ1, we already know that the minor differences in the inputs can be captured by the abstract model. Based on the abstract model, we proposed diverse testing criteria. In this section, we aim to evaluate whether the testing criteria can help to find potential defects and issues in RNNs, i.e., is there some correlation between testing coverage\footnote{In this paper, we focus on the \emph{basic state coverage}, leaving the evaluation of other criteria in the future work.} and erroneous behavior of the RNN? 

We sample 100 audio from the test data whose \emph{Word Error Rates} (WERs) are 0. The WER of the audio represents the error rate for the transcripts from RNN.  
In other words, the sampled audio could be perfectly processed by the RNN.
Based on the 100 audio (seeds), we randomly generate 10,000 audio (with different WER) by metamorphic transformations. 
 
Figure~\ref{fig:correlation} shows the distribution of the 10,000 audio under different abstraction configurations. The $x$-axis is WER and the $y$-axis is the Jaccard Index value. The results show that, in general, test cases with higher relative WER tend to have lower Jaccard index, which means they are less similar with the original seed input.
The more fine-grained the abstract model is, the more obvious is such phenomenon. Intuitively, by increasing the state coverage, we can generate more data which have lower Jaccard index. Thereby attempting to increase the coverage on state offers more possibility to detect more erroneous behaviors of the RNN.

\begin{tcolorbox}[size=title]
{\textbf{Answer to RQ 2:}} There is a strong correlation between the state variation (i.e., the Jaccard Index) and erroneous behaviors of the RNN (i.e., WER). The transformed audio is likely to trigger more erroneous behaviors if it covers more different states comparing with the original audio. By improving BSC coverage, more states are covered and more erroneous behaviors would potentially be captured.
\end{tcolorbox}

\subsection{RQ3: Effectiveness of \tool for coverage improvement}
\label{subsec:rq3}
\tool is designed to generate test cases with high coverage based on the coverage feedback.
We evaluate the effectiveness of \tool in generating high-coverage tests with BSC guidance. 
Experimentally, we sample 100 audio as the initial seeds and run \tool for 12 hours.
Table~\ref{table:rq3:overall} shows the increase of BSC with different abstraction configurations.
Columns ``Ini. Cov.'' and ``12h Cov.'' present the initial coverage of seeds and the coverage 
achieved after 12 hours' testing, respectively.
Column ``Increase'' shows the coverage increases with respect to the initial values. The results show that \tool can increase the coverage effectively. As the configuration becomes more and more fine-grained (from $M_{\{3,10\}}$ to $M_{\{3,100\}}$), the initial coverage is smaller because the states are more in the abstract model. At the same time, the increment of the coverage becomes larger.

\begin{tcolorbox}[size=title]
{\textbf{Answer to RQ 3:}} \tool can obviously improve the state coverage. Furthermore, it is more effective when the abstract model is more fine-grained. 
\end{tcolorbox}

\begin{table}[!t]
\centering
\caption{Basic state coverage increase with \tool.}
\label{table:rq3:overall}
\begin{tabular}{lrr|r}
\toprule
 Config.         & Ini. Cov. & 12h Cov. & Increase \\ \midrule
 $M_{\{3,10\}}$ & 26.1      & 43.2     & 65.5\%   \\
                     $M_{\{3,20\}}$ & 20.1     & 40.5      & 101.5\%     \\
                     $M_{\{3,50\}}$ & 12.3      & 36.2      & 194.3\%    \\
                     $M_{\{3,100\}}$ & 5.5      & 28.4      & 416.4\%    \\
\bottomrule
\end{tabular}
\end{table}

\subsection{RQ4: Erroneous Behavior Detection}
To answer the question, we measure the WER of the generated audio in Fig.~\ref{fig:rq4}. WER represents the erroneous degree of the RNN prediction. Fig.~\ref{fig:rq4} shows the average WER of generated audio under different abstraction configurations. The results show that the average WER is higher for more fine-grained abstract model. It can be explained by the answer to RQ2. With more fine-grained abstract model, \tool will generate test cases that can cover more states of the model. Thus, the test cases are more different with the original seed input (i.e., the Jaccard Index is smaller). Finally, it is more likely to generate test cases with higher WER.  

\begin{tcolorbox}[size=title]
{\textbf{Answer to RQ 4:}} \tool can effectively generate tests to trigger erroneous behaviors of the RNN. For more fine-grained abstract model, it will capture more erroneous behaviors (i.e., higher WER) of the RNN. 
\end{tcolorbox}

\subsection{Threats to Validity}
We list factors which could affect the validity of the experiments.
Due to resource constraints, we did not use all of the training data for constructing the abstract model.
This may result in a abstract model which does not fully reflect the actual trained network behaviors.
To mitigate this problem, we randomly select samples from the training set and follow the distribution of training data approximately. We adopt a conservative metamorphic transformation strategy to make small changes on the original audio such that \tool will generate realistic audio. However, it may still cause false positives especially for the low-quality input (i.e., the audio is not clear such as low volume, too much noise). To mitigate this problem, we manually check and make sure the selected inputs (i.e., the 100 audio in RQ1, RQ2, RQ3 and RQ4) are of high quality.

%% file: relatedwork.tex
\section{Related Work}
\label{sec:related}

\begin{figure}[t]
	\centering
	\includegraphics[width=.5\linewidth]{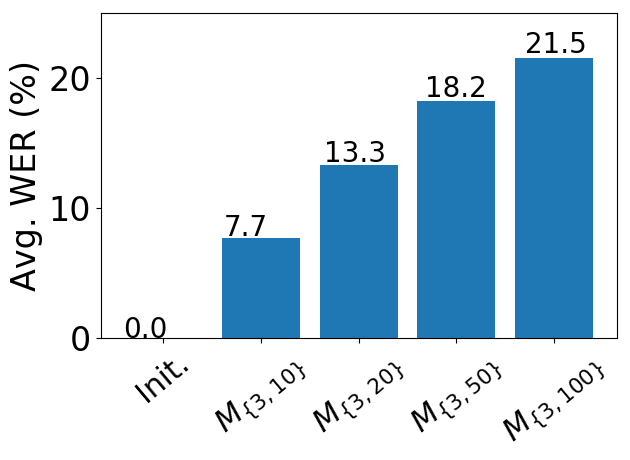}
	\caption{Average WER of inputs generated with \tool.}
	\label{fig:rq4}
\end{figure}

In this section, we compare our work with other testing and abstraction techniques for DL systems.

\paragraph{Testing of DNN}
The lack of robustness places a major threat to the commercialization and wide adoption of DL systems.
Researchers have devoted a great amount of efforts to investigate effective and systematic 
approaches to test DL systems, led by the pioneer work of Pei et al.~\cite{pei2017deepxplore}.
The authors designed the first set of testing criteria  -- \emph{neuron coverage} -- to measure how 
much internal logic of the DNN has been examined by a given set of test data.
Several new criteria have been proposed since then, including a set of adapted MC/DC test criteria 
by Sun et al.~\cite{sun2018testing}, a set of multi-granularity testing criteria by Ma et 
al.~\cite{ma2018deepgauge}, and combinatorial testing criteria~\cite{2018arXiv180607723M}.
So far, the proposed coverage criteria are used to guide the metamorphic mutation-based 
testing~\cite{tian2018deeptest}, concolic testing~\cite{Sun2018}, and coverage-guided testing of DNN~\cite{Odena2018,Xie2018}. In addition, mutation testing techniques are also proposed to evaluate the test data quality through injecting faults into DL models~\cite{8539073}.

The usefulness of MC/DC criteria is limited by the scalability issue, and other criteria are more suitable for the feedforward neural network architecture, even through they are partially applicable to RNN via unrolling.
The experimental results reported in~\cite{tian2018deeptest} demonstrated that the neuron coverage 
works effectively on CNN but far from ideal on RNN when used to guide the generation of test cases.
This indicates that RNN are not simple folding of CNN, and existing criteria may not be well suited for RNN.
Currently, there is still a lack of customized testing criteria specially designed for RNN, which are able to capture 
the statefulness of RNN and measure the thoroughness of the testing efforts.

\paragraph{Abstraction of RNN}
Many approaches have been proposed to model RNN, usually in the form of Finite State Automaton 
(FSA), in order to understand the internal dynamics of RNN.
FSA represents the internal state transitions explicitly and thus can be used to interpret the 
underlying decision rules embedded in RNN.
MDP has similar properties in that sense, and it also captures state transition distributions under 
different inputs, making it a more precise model.
Constructing a FSA from RNN requires two steps: (1) state space partition over the real-valued 
numerical vectors; and (2) automaton construction based on the partitions.
Various partitioning strategies and automaton construction algorithms have been proposed.
Omlin and Giles~\cite{omlin1996extraction} proposed to divide each dimension of the state vector 
into equal intervals, so as to divide the state space into regular grids.
Unsupervised classification algorithms were also applied for state space partitions. 
For example, $k$-means and its variants were studied in
\cite{cechin2003state,Wang:2018:EER:3281485.3281494,Hou2018}.
Weiss et al.~\cite{weiss2017extracting} devised an algorithm to dynamically create partitions, 
where an SVM classifier with an RBF kernel is fitted to separate several state vectors from its 
original partitions.
Recent studies~\cite{weiss2017extracting,Wang:2018:EER:3281485.3281494,Hou2018} have focused more 
on the LSTM and GRU, demonstrating that the same abstraction techniques also work on RNN variants.
When applied to real-world tasks, including natural language processing and speech recognition, the 
state space of the trained RNN model could be tremendously large.
This makes scalability an issue for partition techniques such as $k$-means and kernel algorithms.
Therefore, we adopted the much cheaper interval abstraction and we could also benefit from its 
flexibility in precision adjustment.

%% file: conclusion.tex
\section{Conclusion}
\label{sec:conclusion}

Vulnerabilities of current DL systems, such as autonomous vehicles and voice assistants, are threatening the trust and mass adoption of these technologies.
As such, much research efforts have been focused on the testing of DL systems to ensure their reliability and robustness. 
Yet, little work has been done on the testing of stateful DL systems.
As the first work along this line, we designed a set of test coverage criteria, which can be used to guide the systematic testing of software systems powered by stateful neural networks.

Furthermore, we proposed a set of metamorphic transformations on audios inspired by real-world scenarios, and implemented a general fuzzing framework to discover
defects in ASRs.
We confirmed the usefulness of the proposed criteria on ASRs and showed that the fuzzing framework is effective in exposing real-world defects.
In the future, we plan to evaluate our techniques on more diverse application domains, towards providing quality assurance solution for DL systems life-cycle ~\cite{2018arXiv181004538M}.